\documentclass[11pt]{article}
\usepackage{authblk}
\usepackage{times}
\usepackage{graphicx}
\usepackage{psfig,epsfig}
\usepackage{color}
\usepackage{amsmath}
\usepackage{amssymb}
\usepackage{mathtools}
\usepackage{letterspace}
\usepackage{xspace}
\usepackage[english]{babel}
\usepackage{url}
\usepackage{colortbl}
\usepackage{hhline}
\usepackage{rotating}
\usepackage{relsize}
\usepackage[font={small,it},labelfont=bf]{caption}

\newcommand{\commentout}[1]{}

\DeclareFontFamily{U}{euc}{}% I chose euc because the chart is called Euler cursive
\DeclareFontShape{U}{euc}{m}{n}{<-6>eurm5<6-8>eurm7<8->eurm10}{}%
\DeclareSymbolFont{AMSc}{U}{euc}{m}{n} % I chose AMSc because AMSa and AMSb are defined in the amsfonts-package
\DeclareMathSymbol{\umu}{\mathord}{AMSc}{"16}
\DeclareMathSymbol{\ualpha}{\mathord}{AMSc}{"0B}
\DeclareMathSymbol{\ubeta}{\mathord}{AMSc}{"0C}
\DeclareMathSymbol{\ulambda}{\mathord}{AMSc}{"15}

\newcommand{\sups}[1]{$^{\mathrm{#1}}$}

\usepackage{setspace}
\usepackage{cite}
\title{The relation between alignment covariance and background-averaged epistasis.}

\author[1]{Frank J. Poelwijk \thanks{Email: \texttt{poelwijk@gmail.com}}}

\author[2]{Rama Ranganathan}
\affil[1]{cBio Center, Dana-Farber Cancer Institute, 360 Longwood Avenue, Boston, Massachusetts, 02115, USA}
\affil[2]{Green Center for Systems Biology and Departments of Biophysics and Pharmacology, UT Southwestern Medical Center, 6001 Forest Park Road, Dallas, Texas, 75235, USA}
\date{}
\begin{document}
\maketitle

Epistasis, or the context-dependence of the effects of mutations, limits our ability to predict the functional impact of combinations of mutations, and ultimately our ability to predict evolutionary trajectories. Information about the context-dependence of mutations can essentially be obtained in two ways: First, by experimental measurement the functional effects of combinations of mutations and calculating the epistatic contributions directly, and second, by statistical analysis of the frequencies and co-occurrences of protein residues in a multiple sequence alignment of protein homologs. In this manuscript, we derive the mathematical relationship between epistasis calculated on the basis of functional measurements, and the covariance calculated from a multiple sequence alignment. There is no one-to-one mapping between covariance and epistatic terms: covariance implies epistasis, but epistasis does not necessarily lead to covariance, indicating that covariance in itself is not the directly relevant quantity for functional prediction. Having calculated epistatic contributions from the alignment, we can directly obtain a functional prediction from the alignment statistics by applying a Walsh-Hadamard transform, fully analogous to the transformation that reconstructs functional data from measured epistatic contributions. This embedding into the Hadamard framework is directly relevant for solidifying our theoretical understanding of statistical methods that predict function and three-dimensional structure from natural alignments.

\subsection*{Introduction}

Some time ago the traditional scope of epistasis has been extended from interactions between pairs of mutations, to interactions between triple, quadruple, and larger sets of mutations (see e.g. \cite{JMolBiol224_733}). An effective way to characterize this \emph{so-called} higher-order epistasis is by applying a Walsh-Hadamard transform to experimentally obtained phenotypic data \cite{BiolCybern65_321,CurrOpinGenetDev23_700,PLoSComputBiol12_e1004771}. In this framework, the epistatic contributions $\boldsymbol{\bar{\omega}}$ for all combinations of mutations are obtained by performing a linear transformation denoted as $\boldsymbol{\bar{\omega}} = \boldsymbol{V} \boldsymbol{H} \boldsymbol{\bar{y}}$ \cite{PLoSComputBiol12_e1004771}. Here $\boldsymbol{\bar{y}}$ is a vector of the phenotypic data for a complete combinatorial mutant dataset, i.e. measured phenotypes for all combinations of amino acid mutations between two initial sequences. $\boldsymbol{H}$ is the Walsh-Hadamard matrix \cite{AmJPhys49_466}, and $\boldsymbol{V}$ is a weight matrix that provides pre-factors for the terms. Conversely, since this is a one-to-one mapping, if we know all epistatic contributions between mutations, we are able to make reconstruct all phenotypes by applying the the inverse transform ($\boldsymbol{\bar{y}}  = \boldsymbol{H}^{-1} \boldsymbol{V}^{-1} \boldsymbol{\bar{\omega}}$). This particular of epistasis was termed 'background-averaged epistasis', indicating that, for example, a first-order contribution is the effect of a mutation at one position, averaged over all combinations of mutations at the remaining positions. This distinguishes the description from the more traditional, or local, epistasis, where all effects are calculated with respect to a chosen reference sequence.

A complementary approach to investigate mutational context-dependence is by statistical analysis of the occurrences and co-occurrences of amino acids in multiple sequence alignments of protein homologs. The reasoning behind this is that if some residue is beneficial globally across the homologs, it will occur in a multiple sequence alignment with a frequency higher than the background expectation. Similarly, pairs of residues that are beneficial, or deleterious, should be over- or under-represented in a representative alignment. Methods have been developed to extract various types of information, from cooperative interactions between amino acids \cite{Cell138_774}, to three-dimensional structure \cite{LapedesLANL2002,NatBiotechnol30_1072}, and protein-protein interactions\cite{PNAS106_67}. Additionally, global probability models have been applied to the prediction of phenotypes in mutant libraries \cite{NatBiotechnol35_128}.

In the current manuscript we directly calculate epistatic terms from an alignment of functional sequences and compare this to covariance terms based on amino acid co-occurrences. To do this, we initially make the assumption that the phenotype is categorical, i.e. assumes the values 0 and 1 for non-functional and functional sequences, which allows us to directly compare epistasis and covariance terms. We argue that the categorical assumption will not affect the applicability of the theoretical framework, as sequences in natural alignments can be reasonably expected to be highly functional. We use experimental data from \cite{GFPpaper}, where a combinatorially complete dataset consisting of 2\sups{13} mutants in the \emph{Entacmaea quadricolor} fluorescent protein was phenotypically assayed for color and brightness, to show that the assumption is justified. Having distilled epistatic contributions from an alignment, we can now make functional predictions fully analogous to the procedure in the background-averaged epistasis framework, by inversion of the transformation $\boldsymbol{\hat{y}}^\mathrm{aln} = \boldsymbol{H^{-1}} \boldsymbol{V^{-1}} \boldsymbol{\bar{\omega}}^\mathrm{aln}$, where $\boldsymbol{\hat{y}}^\mathrm{aln}$ designates that functional prediction.

\subsection*{Results}
For a certain combinatorial sequence space, if we were able to make an alignment that contained all functional sequences in that space, we would write for the frequency $f_i$ of the amino acid designated by '1' at position $g_i$
\begin{equation}
f_i = \frac{N_{g_i=1}^\mathrm{func}}{N_\mathrm{tot}^\mathrm{func}},
\label{eq:deffreq}
\end{equation}
where $N_{g_i=1}^\mathrm{func}$ is the number of sequences in the alignment with amino acid '1' at position $g_i$, and $N_\mathrm{tot}^\mathrm{func}$ represents the total number of functional sequences. We can write for the first-order background-averaged epistatic term $\boldsymbol{\omega}_i$ associated with the same position
\begin{align}
\boldsymbol{\omega}_i & = \frac{1}{N_\mathrm{tot}/2}\ \big(\, y_{10000...}-y_{00000...}+y_{11000...}-y_{01000...} + ...\  \big)
\end{align}
where $N_\mathrm{tot}$ is the total size of the sequence space and $y_{...}$ denote the phenotypes associated with the subscripted genotypes. The signs in front of the phenotypes correspond to the signs of the row in the Hadamard matrix that calculates the first-order background-averaged epistatic contribution for position $i$ (see \cite{PLoSComputBiol12_e1004771}). If we now make the assumption that all phenotypes are either functional, '1', or nonfunctional, '0', we can write for the first-order background-averaged epistatic terms \cite{PLoSComputBiol12_e1004771}
\begin{figure}[t]
\center \psfig{figure=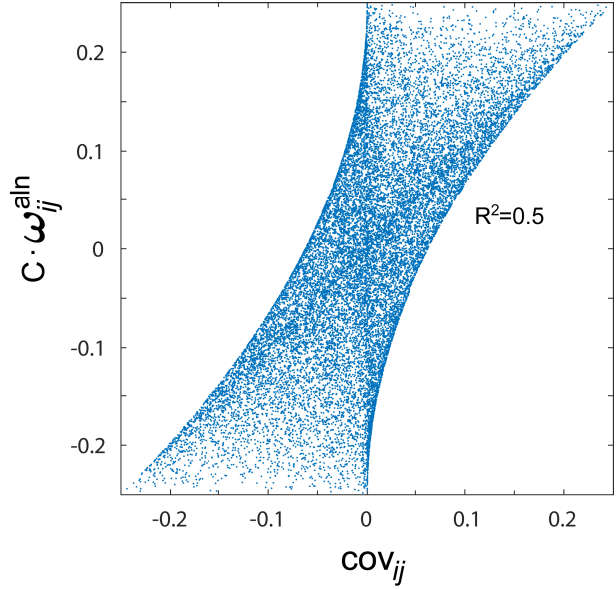,width=.5\linewidth}
\caption{Second-order alignment epistatic terms versus covariance. Shown are the results of a simulation in which both the covariance $f_{ij} - f_i f_j$ and the second-order alignment epistatic terms $\propto{}4f_{ij} - 2(f_i+f_j) + 1$, were calculated for 20,000 random choices of $f_i$ and $f_j$, and random compatible choices for $f_{ij}$. As can be seen, there is no one-to-one relation, only a correlation with a value of 0.5. Large (absolute) values for covariance imply large values for alignment epistasis, but the reverse is not true.}
\label{fig:covbHbin}
\end{figure}
\begin{align}
\boldsymbol{\omega}^\mathrm{aln}_i & = \frac{N_{g_i=1}^\mathrm{func}-N_{g_i=0}^\mathrm{func}}{N_\mathrm{tot}/2} = \frac{2N_{g_i=1}^\mathrm{func}-N_\mathrm{tot}^\mathrm{func}}{N_\mathrm{tot}/2} \notag\\[1em]
& = (2 f_i - 1)\ \frac{2 N_\mathrm{tot}^\mathrm{func}}{N_\mathrm{tot}}
\label{eq:omegafreq}
\end{align}
where we obtain a direct link to the amino acid frequencies $f_i$ in the alignment. In what follows we will designate such epistatic terms obtained from the alignment by 'alignment epistais', $\boldsymbol{\bar{\omega}}^\mathrm{aln}$.\\
\\
Now we can find the second-order terms in a similar fashion
\begin{equation}
f_{ij} = \frac{N_{g_{ij}=11}^\mathrm{func}}{N_\mathrm{tot}^\mathrm{func}}, \ \ \ \ f_i = \frac{N_{g_{ij}=11}^\mathrm{func}+N_{g_{ij}=10}^\mathrm{func}}{N_\mathrm{tot}^\mathrm{func}}, \ \ \ \ f_j = \frac{N_{g_{ij}=11}^\mathrm{func}+N_{g_{ij}=01}^\mathrm{func}}{N_\mathrm{tot}^\mathrm{func}}
\end{equation}
allowing us to express the second-order background-averaged epistatic terms in terms of frequencies
\begin{align}
\boldsymbol{\omega}^\mathrm{aln}_{ij} & = \frac{1}{N_\mathrm{tot}/4} \big(\,N_{g_{ij}=11}^\mathrm{func}-N_{g_{ij}=10}^\mathrm{func}-N_{g_{ij}=01}^\mathrm{func}+N_{g_{ij}=00}^\mathrm{func}\ \big) \notag \\[1em]
& = \big(\, 4 f_{ij} - 2(f_i+f_j) + 1 \ \big)\ \frac{4 N_\mathrm{tot}^\mathrm{func}}{N_\mathrm{tot}}
\end{align}
This clarifies the similarity to the covariance of frequencies in the alignment defined by
\begin{equation}
\mathrm{cov}_{ij} = f_{ij} - f_i f_j
\end{equation}
since $\boldsymbol{\omega}^\mathrm{aln}_{ij}$ and $\mathrm{cov}_{ij}$ share the same leading term $f_{ij}$. To be more precise, the relation between $\boldsymbol{\omega}^\mathrm{aln}_{ij}$ and $\mathrm{cov}_{ij}$ cannot be written down as a closed form expression. We can see from a simple simulation based on random values for $f_i$, $f_j$, and compatible choices for $f_{ij}$ that covariance implies epistasis, but that epistasis does not imply covariance (Fig \ref{fig:covbHbin}). There is a general relationship between the two quantities with an R\sups{2} of 0.50.\\

\begin{figure}[t]
\center \psfig{figure=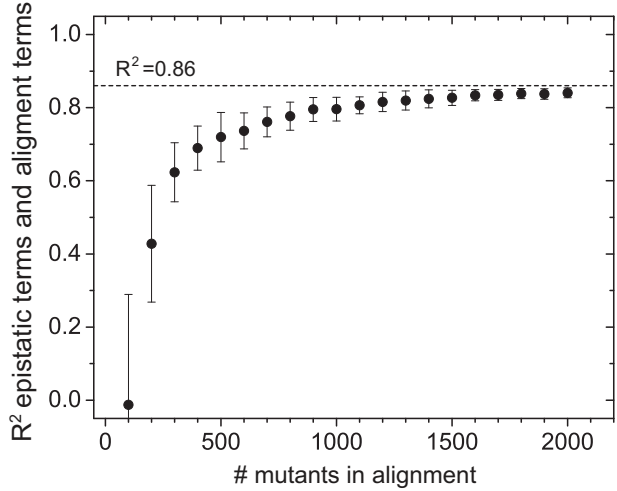,width=.5\linewidth}
\caption{R\sups{2} between 1st and 2nd order background-averaged epistatic versus alignment terms. The background-averaged terms were obtained based on the functional values for the complete dataset \cite{GFPpaper}, according to the transform $\boldsymbol{\bar{\omega}} = \boldsymbol{V} \boldsymbol{H} \bar{\mathcal{E}}$ \cite{PLoSComputBiol12_e1004771}. Shown here is the goodness-of-fit (R\sups{2}) between first- and second-order terms from $\boldsymbol{\bar{\omega}}$ versus $\boldsymbol{\bar{\omega}}^\mathrm{aln}$ as a function of an increasing number of functional sequences in the alignment that is used to calculate the alignment epistasis $\boldsymbol{\bar{\omega}}^\mathrm{aln}$. When all functional mutants in the combinatorial space are included (2984 out of 8192 total mutants) R\sups{2} reaches 0.86.}
\label{fig:GoFbHbin12}
\end{figure}
Clearly, some initial idealizations were made to arrive at this result. First, no realistic alignment will contain \emph{all} functional sequences. However, since the relevant terms are frequencies, they should be robust if the alignment is a representative subset of the functional sequences. Indeed, Fig. \ref{fig:GoFbHbin12}, where we calculated the goodness-of-fit between epistatic terms based on experimental data in \cite{GFPpaper} and alignment epistasis, illustrates that the two rapidly converge for alignments of limited depth. The second assumption, that of binary phenotypes, may be quite reasonable for sequences that are found in nature, because we generally observe the outcomes of evolutionary optimization instead of intermediates. However, even for the mutational dataset in \cite{GFPpaper} where we do observed a range of phenotypic values, the similarity between alignment epistasis and background-averaged epistasis from the functional assay is surprisingly good (when we include all functional mutants, the value for R\sups{2} for first and second-order terms combined is 0.86).\\

In fact, we can derive an expression to calculate for every order of alignment epistatic terms, $\boldsymbol{\bar{\omega}}^\mathrm{aln}$. In the ideal, but unrealistic, case of the alignment containing all functional sequences, this constitutes an invertible decomposition. In the realistic case where the alignment only contains a fraction of all functional sequences, one will have to determine how many orders of alignment epistasis can be meaningfully calculated before noise takes over.\\

\begin{figure}[t]
\center \psfig{figure=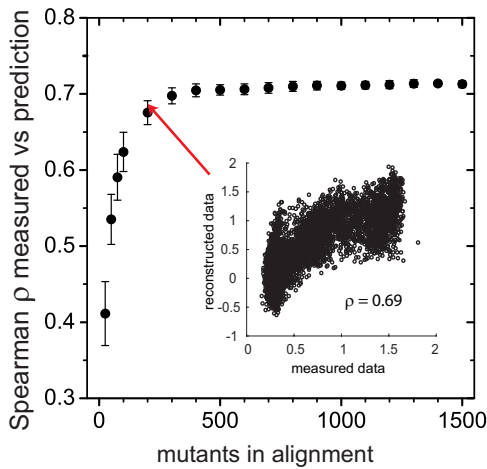,width=.4\linewidth}
\caption{Spearman correlation coefficient $\rho$ calculated between predicted phenotypic values from alignment epistasis and actual measurements. A high $\rho$ is reached for limited alignment depths. The inset shows a typical reconstruction based on an alignment of 200 sequences (of 8192 total).}
\label{fig:Spearman}
\end{figure}

The current demonstration becomes more straightforward if we reparametrize the genotype as $\sigma_i = -1/1$ instead of $g_i = 0/1$ (in ref \cite{PLoSComputBiol12_e1004771} we showed that this reparametrization naturally leads to a background-averaged description). Now we can write, in analogy to the frequency $f_i$ in equation \ref{eq:deffreq}, for the average value of a column in the alignment
\begin{equation}
\phi_{i} = \frac{N_{\sigma_i=1}^\mathrm{func}-N_{\sigma_i=-1}^\mathrm{func}}{N_\mathrm{tot}^\mathrm{func}}
\end{equation}
Comparison to equation \ref{eq:omegafreq} identifies $\phi_i$ with $2f_i - 1$ and
\begin{equation}
\boldsymbol{\omega}^\mathrm{aln}_i = \phi_i\ \frac{2 N_\mathrm{tot}^\mathrm{func}}{N_\mathrm{tot}}
\end{equation}
Now it is straightforward to show that
\begin{equation}
\boldsymbol{\omega}^\mathrm{aln}_{ij} = \phi_{ij} \frac{4 N_\mathrm{tot}^\mathrm{func}}{N_\mathrm{tot}},\ \ \ \ \mathrm{and}\ \ \ \ \boldsymbol{\omega}^\mathrm{aln}_{ijk} = \phi_{ijk} \frac{8 N_\mathrm{tot}^\mathrm{func}}{N_\mathrm{tot}},\  \mathrm{etc.}
\end{equation}
Computationally, the $\phi_{ij...yz}$ are simply obtained by taking the mean value of the element-wise product of entries in columns $ij...yz$.
The obtained vector $\boldsymbol{\bar{\omega}}^\mathrm{aln}$ now can be used in exactly the same way as the regular background-averaged epistasis vector $\boldsymbol{\bar{\omega}}$ to calculate the phenotypic values for all the mutants in the combinatorial space
\begin{equation}
\boldsymbol{\hat{y}}^\mathrm{aln} = \boldsymbol{H^{-1}} \boldsymbol{V^{-1}} \boldsymbol{\bar{\omega}}^\mathrm{aln}
\end{equation}
Doing this for the data in \cite{GFPpaper}, we see that reasonable Spearman correlations are reached even for very limited alignment depths; in this case containing 200-300 of more than 8000 sequences (Fig. \ref{fig:Spearman}).

\subsection*{Discussion}
The prediction of the functional effects of multiple mutations has been the focus of several recent studies, either based on high-throughput experimental functional characterization (e.g. \cite{NatureGenetics43_487}), or based on amino acid frequency statistics in natural sequence alignments (e.g. \cite{NatBiotechnol35_128}). The latter approach has originated from the 3D structure prediction community using global probability models for the sequences \cite{NatBiotechnol30_1072}. In the current work we derived the link between amino acid covariance and background-averaged epistasis \cite{PLoSComputBiol12_e1004771} from first principles. We showed that covariance implies epistasis, but that the converse is not true. We subsequently showed that the obtained terms can be used in a straightforward way, using an inverse Hadamard transform, to make a functional prediction for all mutants in the combinatorial space. Making the identification between experimentally obtained epistasis and and alignment statistics should help us solidify the theoretical basis for covariance methods that predict function and 3D structure from natural alignments. That said, completing the theoretical description, and ensuring its direct applicability to natural libraries -where each position has 20 amino acids choices instead of 2- will require an explicit extension of the Hadamard framework to a cardinality larger than 2, which, as of yet, has not been made.

\end{document}